\documentclass[fleqn,usenatbib]{mnras}

\usepackage{newtxtext,newtxmath}

\usepackage[T1]{fontenc}
\usepackage[normalem]{ulem}

\DeclareRobustCommand{\VAN}[3]{#2}
\let\VANthebibliography\thebibliography
\def\thebibliography{\DeclareRobustCommand{\VAN}[3]{##3}\VANthebibliography}

\usepackage{graphicx}	
\usepackage{amsmath}	
\usepackage{gensymb}
\usepackage{wrapfig}
\usepackage{float}
\usepackage{caption}
\usepackage{subcaption}
\usepackage{xcolor}
\usepackage{soul}
\usepackage[dvipsnames]{xcolor}
\usepackage[subtle]{savetrees}

\title[Fast jets in Scorpius X-1]{Exploring the potential for ultra-relativistic jets in Scorpius X-1 with low angular resolution radio observations}

\author[I. Stephens et al.]{
I. Stephens$^{1}$\thanks{E-mail: stephensisabel0@gmail.com},
L. Rhodes$^{2,3}$\thanks{E-mail: lauren.rhodes@mcgill.ca},
A.J. Cooper$^{4}$,
S.E. Motta$^{5}$,
J.S. Bright$^{4}$
\\
$^{1}$School of Physics and Astronomy, University of Leeds, Leeds, LS2 9JT, UK\\
$^{2}$Trottier Space Institute at McGill, 3550 Rue University, Montreal, Quebec H3A 2A7, Canada\\
$^{3}$Department of Physics, McGill University, 3600 Rue University, Montreal, Quebec H3A 2T8, Canada\\
$^{4}$Department of Physics, Astrophysics, University of Oxford, Denys Wilkinson Building, Keble Road, Oxford, OX1 3RH, UK\\
$^{5}$INAF-Osservatorio Astronomico di Brera, Via Bianchi 46, I-23807, Merate, I-23807 (LC) Italy\\
}

\date{Accepted XXX. Received YYY; in original form ZZZ}

\pubyear{\the\year{}}

\begin{document}
\label{firstpage}
\pagerange{\pageref{firstpage}--\pageref{lastpage}}
\maketitle

\begin{abstract}
Scorpius X-1 (Sco X-1) is a neutron star X-ray binary in which the neutron star is accreting rapidly from a low mass stellar companion. At radio frequencies, Sco X-1 is highly luminous and has been observed to have jet ejecta moving at mildly relativistic velocities away from a radio core, which corresponds to the binary position. In this Letter, we present new radio observations of Sco X-1 taken with the Karl G. Jansky Very Large Array. Using a fast imaging method, we find that the 10 and 15\,GHz data show a number of flares. We interpret these flares as the possible launching of fast jets ($\beta\Gamma>2$), previously observed in Sco X-1 and called ultra-relativistic flows, and their interaction with slower moving jet ejecta. Using the period between successive flares, we find that it is possible for the fast jets to remain undetected, as a result of the fast jet velocity being sufficiently high to cause the jet emission to be beamed in the direction of the motion and out of our line of sight. Our findings demonstrate that the ultra-relativistic flows could be explained by the presence of fast jets in the Sco X-1 system. 

\end{abstract}




\section{Introduction}\label{sec:intro}
Relativistic outflows, resulting from accretion onto a compact object, are a common feature of X-ray binaries. Such relativistic outflows, also known as jets, can be observed in `steady' and `transient' forms. The steady form is a compact, self-absorbed outflow thought to be mildly relativistic, with speeds likely spanning $\lesssim$0.5c -- 0.9c \citep{2009MNRAS.396.1370F, Tetarenko2019, Tetarenko2021}, while the transient form can range from $\sim$0.1c to highly relativistic velocities, including observed superluminal motion \citep[e.g.][]{1999ARA&A..37..409M, 2006csxs.book..381F, Bright2020, Carotenuto2021, Bahramian2023}. The observed radio emission from jets may be a result of jet interactions with the ambient medium material, causing a relativistic shock, or in situ particle acceleration \citep{2025arXiv250310804C}.  

Scorpius X-1 (Sco X-1) is a well-known low-mass X-ray binary system hosting a neutron star, and is considered a prototypical Z-source. These are systems accreting at very high rates which buries the neutron star's magnetic field and produces luminous X-ray emission which traces a `Z' shape on an X-ray colour-colour diagram \citep{1989A&A...225...79H,2006csxs.book..381F}.  

\begin{figure*}
    \centering
    \includegraphics[width=\textwidth]{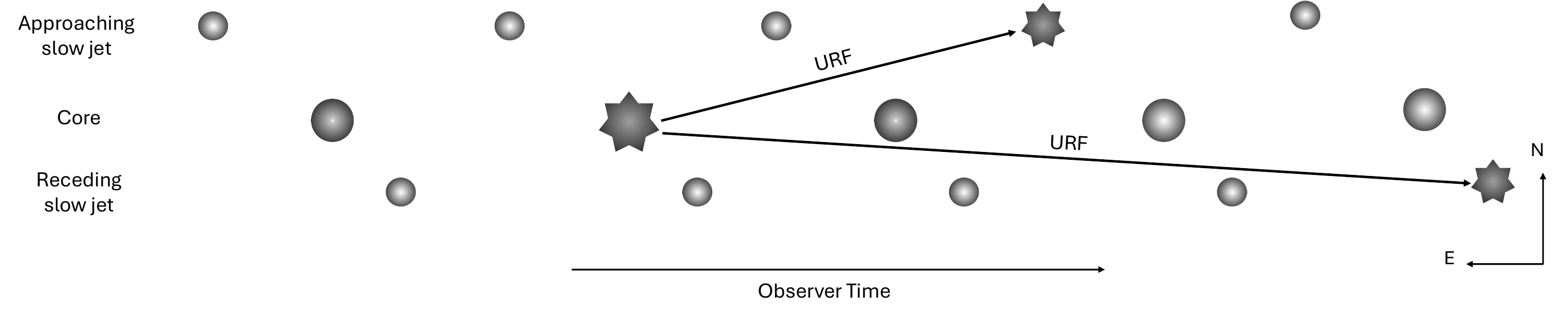}
    \caption{A schematic of Sco X-1 showing the positions relative positions of the core, approaching and receding slow jets and the effects of the URF on the system. The URF (indicated by the diagonal arrows) is launched from the core at the time of the core's flare and then propagates out to the slow jets where they then cause flares, first in the approaching and receding jet.}
    \label{fig:schematic}
\end{figure*}

Sco X-1 also produces bright radio emission that originates from multiple components, including a core, corresponding to the binary position, and two lobes, radio-bright jet ejecta slowly moving away from the launch site. Figure \ref{fig:schematic} presents a schematic of the Sco X-1 system. The approaching jet (the top left blob in each panel of Figure \ref{fig:schematic}) is observed as a lobe moving to the north-east (NE), and the receding jet is observed as a fainter lobe (the bottom right blob in each panel of Figure \ref{fig:schematic}), visible less often (half of the time) moving to the south-west \citep[SW, ][]{2001ApJ...558..283F, 2019MNRAS.483.3686M}. The radio lobes of Sco X-1 have been seen to travel at $\sim$0.5c and have been spatially resolved on milli-arcsecond scales \citep{2001ApJ...558..283F}. 

Highly relativistic and even superluminal jets have been observed in both black hole and neutron star binaries \citep[e.g.][]{1994Natur.371...46M, 2020ApJ...891L..29H}. It has been proposed that there may exist a highly relativistic and unseen flow of energy interacts with the observed radio jets, called ultra-relativistic flow of energy (URF), in neutron star binaries \citep[the authors note that URFs have only been observed in neutron star binaries][]{2004Natur.427..222F, 2019MNRAS.483.3686M}. In \citet{2001ApJ...558..283F}'s study of the radio properties of Sco X-1, using VLBI (Very Long Baseline Inteferometry) data, confirmed by \citet{2019MNRAS.483.3686M}, they inferred the presence of such unseen URFs, which caused a correlated pattern of flares in the flux density of the radio core and lobes. A burst of energy at the core, propagating down the path of the existing radio jet (the lobe or `slow' jet), could intercept both lobes, causing them to re-brighten. Due to the viewing angle and relativistic effects, the flares would always be observed in the order of the core first, then the approaching lobe, then the receding lobe. The URF was not directly detected in either the radio or X-ray data, but its presence could only be inferred based on the flaring behaviour, and were inferred to be highly relativistic with speeds of $\approx0.95$c \citep{2001ApJ...553L..27F}. For ease of understanding the difference between the different jets and outflows, we present a summary in Table \ref{tab:jets}.

\begin{table}
\centering
\setlength{\tabcolsep}{3pt} 
\renewcommand{\arraystretch}{1.2} 
\begin{tabular}{p{2cm}p{6.3cm}} 
\hline
Type & Description \\
\hline
Slow jet & Also referred to as radio lobes. These move at velocities of $<0.7c$ and are directly observed in neutron star and black hole X-ray binaries. These are observed as spatially resolved blobs with respect to the core of the binary in Sco X-1. \\
Fast jet & Jets with velocities $>0.7c$. These have only been observed in black hole X-ray binaries. In this work, we are exploring whether fast jets could be present in the Sco X-1 system.  \\
URF & An inferred flow of energy with velocities $>0.95c$. These have only been observed in neutron star X-ray binaries. URFs have been inferred in the Sco X-1 system \\
\hline
\end{tabular}
\caption{A summary of the classification of jets in X-ray binaries \citep{2001ApJ...553L..27F, Fender_Motta_2025}.}
\label{tab:jets}
\end{table}

In this work, we analyse new radio observations of Sco X-1 taken in 2019, using the Karl G. Jansky Very Large Array in the X and Ku bands (9 and 15\,GHz, respectively). The new observations, described in Section \ref{sec:obs} cover a six hour period and have a high temporal resolution, allowing for the timing of flares to be tracked with a higher accuracy than previous observations. In Section \ref{sec:analysis}, we analyse whether flares in the radio spectrum of Sco X-1 could be explained by the presence of a URF, whether flux beaming could allow a URF to remain unseen, and whether URFs could be relativistic `fast' jets as presented in the \citet{Fender_Motta_2025} framework. It is possible that URFs are intrinsically dark, or that the emission is beamed out of our line of sight. In this work, we attempt to constrain the URF properties without VLBI observations, relying on the assumption that URFs are fast jets that may be beamed out of the line of sight, and that the properties of URFs and lobes (slow jets) are very similar.

\section{Observations}\label{sec:obs}

As part of a larger multi-wavelength campaign Sco X-1 was observed with the Karl G. Jansky Very Large Array (VLA) beginning on 24 February 2019 under project code 19A-302 (PI Motta). The VLA was in C->B configuration and was sub-arrayed into three independent arrays observing between UT 08:59:50 and UT 17:00:10. Two of the sub-arrays used the Ku-band (central frequency $15$\,GHz with $2$\,GHz bandwidth and 8-bit sampling) receiver and had offset start times such that target was always being observed. The third sub-array observed at X-band (central frequency 9\,GHz with $2$\,GHz bandwidth and 8-bit sampling). For both frequency setups the bright and unresolved source J1558$-$1409 was used as an interleaved complex gain calibrator (and for pointing calibration at Ku-band), and 3C286 was observed to set the absolute flux density scale and calibrate the bandpass response of each sub-array.

The data from each sub-array was iteratively calibrated using the VLA CASA (version 6.4.1.12; \citealt{2007ASPC..376..127M}) pipeline (version 2022.2.0.64). After each run the pipeline weblog was inspected to check the quality of the calibration, and manual flagging followed by a rerun of the pipeline (with the Hanning smoothing step removed). We particularly note the need for additional flagging at the beginning of some complex gain calibrator scans where the array had clearly not settled onto the target.

Once the data from each sub-array were phase reference calibrated we made deep images from each sub-array using the CASA task \texttt{tclean} with a Briggs robust parameter of $0.5$ for X-band and $-0.5$ for Ku-band, two Taylor terms to describe the spectra of clean components, and with wprojection enabled (and the number of planes automatically calculated by CASA). Clean regions were defined interactively and data were averaged in time and frequency according to the constraints defined in EVLA Memo 199\footnote{\url{https://library.nrao.edu/public/memos/evla/EVLAM_199.pdf}} to reduce processing times (channel width from $1$\,MHz to $8$\,MHz for the Ku-band sub-arrays, and from $1$\,MHz to $4$\,MHz with the X-band sub-array, and time averaging from $2\,\rm{s}$ seconds to $8\,\rm{s}$ for Ku-band and from $4\,\rm{s}$ seconds to $8\,\rm{s}$ for X-band). Clear artefacts remained around Sco X-1 and two other field sources in these deep images.

To improve the quality of the data calibration we performed phase-only self-calibration per sub-array, finding significant improvements in all three cases. For the Ku-band sub-arrays a single iteration of phase-only self-calibration with a solution interval of 1 hour was sufficient to significantly improve the deep images, whereas X-band required two rounds with solution intervals of thirty minutes followed by the scan length. Calibration solutions for each polarisation were calculated independently and on a per-spectral window basis. The phase solutions for each polarisation were not consistently similar, leading us to apply separate solutions. After initial rounds of long-interval self-calibration, additional rounds were attempted but a large number of failed solutions as well as noise-like phase solutions (oscillating without much structure around 0 phase) resulted in no further applications of self-calibration. The lack of further self-calibration rounds was due to the relative low flux density of additional sources in the field and the use of sub-arrays, reducing the sensitivity of the instrument. We did not employ amplitude self-calibration as we wished to measure the short term variability of Sco X-1, which amplitude self-calibration would alter artificially. 

Once the data from each sub-array was self-calibrated we proceeded with imaging on short timescales. For each frequency (after concatenating the data from the two Ku-band sub-arrays) we separated the self-calibrated measurement set into individual measurement sets each covering 60 seconds. Each 60 second section was imaged using natural weighting to ensure maximum sensitivity using the clean masks defined during the creation of our deep images, with a stopping threshold set to twice the image RMS noise of the dirty image. The \texttt{tclean} task was run twice with an updated threshold in the second run using the RMS noise from the first clean image. Due to the poor \textit{uv}-coverage achieved for a 60 second sub-arrayed observation (which results in a particularly poor psf when imaging using natural weighting), we set the \texttt{psfcutoff} parameter to 0.8 to ensure that only the central lobe of the dirty beam was fit when defining the synthesised beam. Sco X-1 is clearly detected in each 60 second image at both frequencies. Data were not combined over scan boundaries and so in some cases a data point represents less than 60 seconds of data. To extract the flux from each image we use the CASA task \texttt{imfit}, with the source size fixed to that of the synthesised beam for a particular observation and fitting within a small region centered on Sco X-1. We add a 5\% systematic flux scale error in quadrature with the error returned by \texttt{imfit}. 

We observe clear correlated variability on short timescales at both X- and Ku-band. Comparison to a field source, which was constant over for the length of the observation, demonstrated that the variability associated with Sco X-1 is real. The light curves are shown in Figure \ref{fig:lightcurve_both} and the data used to create the figure are available as a machine readable table as part of the online version of this article.

\section{Analysis}\label{sec:analysis}
For ease of comprehension, we have divided the analysis into three parts. We first determine the proper motions of the radio lobes of Sco X-1, using digitised data from \citet{2001ApJ...553L..27F, 2001ApJ...558..283F} and verifying the results of \citet{2019MNRAS.483.3686M}. The verification of \citet{2019MNRAS.483.3686M}'s work is necessary as the results are directly used in all subsequent analysis. Secondly, we analyse the 2019 VLA observations of Sco X-1's flares to infer the speeds of URFs, assuming URFs were responsible for sequences of flares and had properties as described in Section \ref{sec:intro}. Finally, we explore whether the relativistic beaming of the inferred URFs could result in them being undetectable in the 2019 observations.

\subsection{Slow Jet Analysis}
\label{subsec:slowjet}

\begin{figure}
    \centering
    \begin{subfigure}[b]{\linewidth}
        \includegraphics[width=\linewidth]{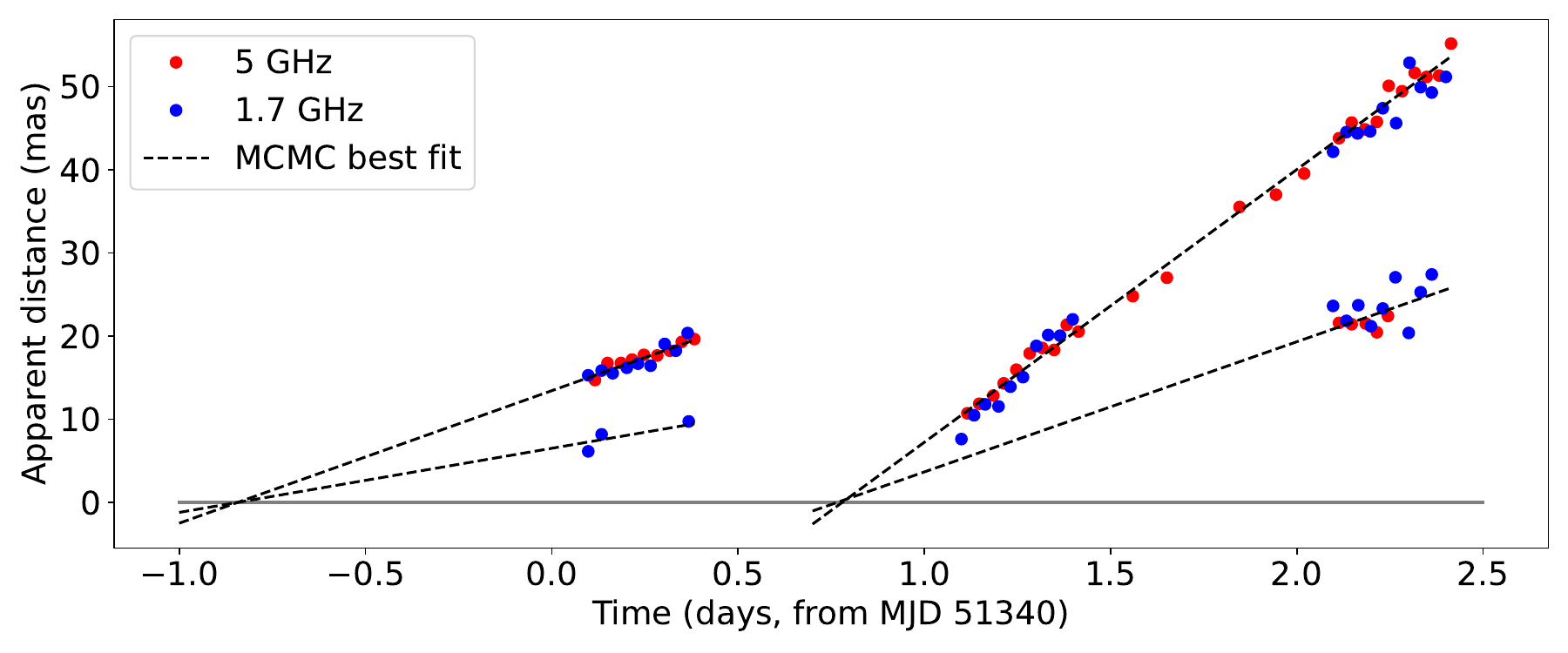}
    \end{subfigure}
    \begin{subfigure}[b]{\linewidth}
        \includegraphics[width=\linewidth]{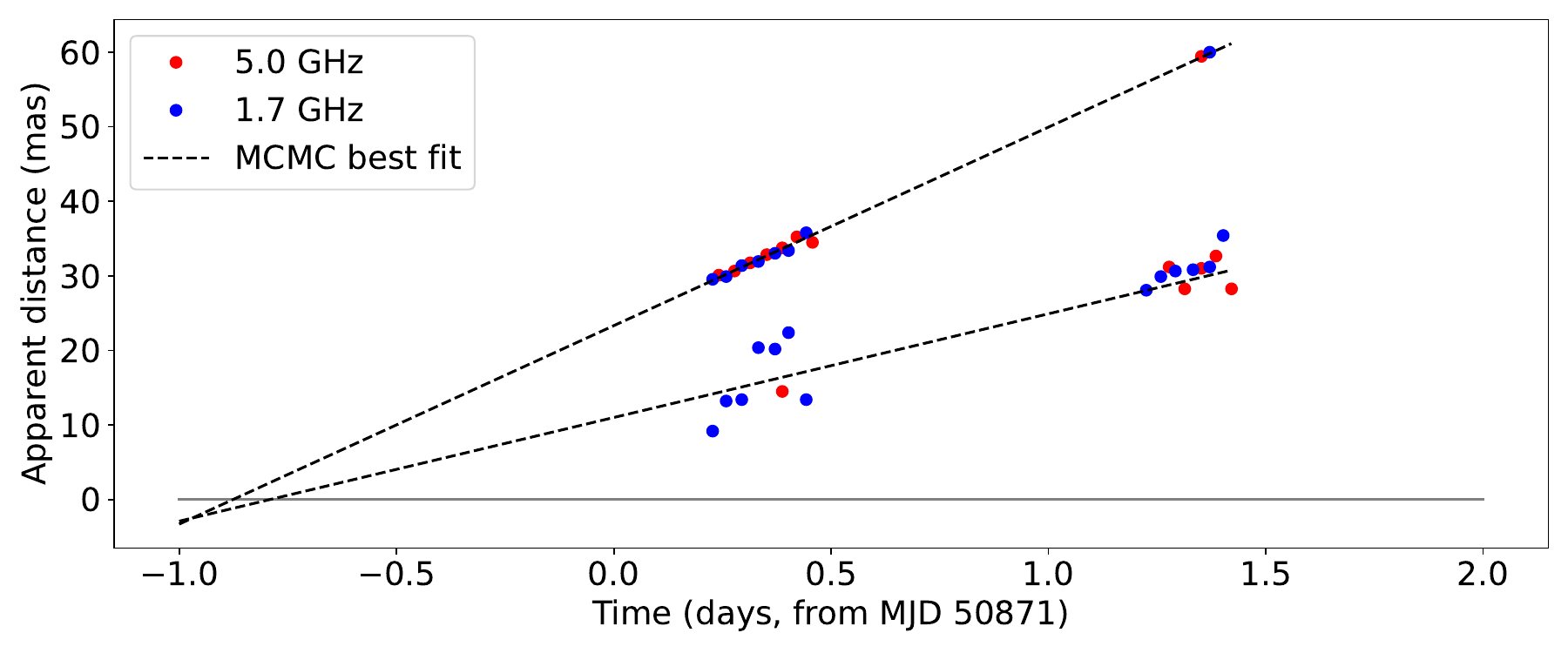}
    \end{subfigure}
    \caption{Two plots showing the observed separation, in mas, of three pairs of radio lobes from the core of Sco X-1, as a function of time. Overlaid is a linear fit to data at both frequencies which corresponds to proper motion in mas/day. We assumed the proper motion was constant. \textit{Upper panel}: Pairs 1 and 2, observed in June 1999. \textit{Lower panel} Pair 3, observed in February 1998.}
    \label{fig:lobes_motion}
\end{figure}

We used all of the digitised data from figure 7 of \citet{2001ApJ...558..283F}, showing the angular separation with time of three pairs of lobes/slow jets. Lobe pairs 1 and 2 were observed in June 1999, and lobe pair 3 was observed in February 1998. Figure \ref{fig:lobes_motion} shows this data, along with linear fits corresponding to the proper motion. We first used linear regression (\textsc{scipy}) to determine the line of best fit parameters and then used these values as the flat priors for a Monte Carlo Markov Chain (\textsc{mcmc}) method \citep[\textsc{emcee}, ][]{2013PASP..125..306F}. The \textsc{mcmc} sampler consisted of 50 walkers and ran for 1000 steps. Convergence was checked for by eye in the trace plots and determined that the first 100 steps needed to be discarded. We obtained posterior distributions for the gradient and y-intercept which we used in our further analysis (Sections \ref{subsec:URF_vla} and \ref{subsec:beaming}).

We assume that the NE and SW lobes are launched at the same time and infer the ejection times by extrapolating backwards to a separation of 0 milliarcseconds (mas). The ejection times quoted in Table \ref{tab:motions} consider only extrapolate of the NE lobe, due to better data quality, however in all cases the inferred ejection time for the SW lobe was in agreement with that of the NE lobe. For the fit of the second NE lobe, we considered only the 5 GHz data, as in \citet{2019MNRAS.483.3686M}, due to better data quality and for consistency. 

\def\arraystretch{1.2}
\begin{table*}
    \begin{tabular}{lccccccc} 
    \hline
      & Ejection time (MJD) & $\mu_{app}$ (mas/day) & $\mu_{rec}$ (mas/day) & $\beta cos\theta$ & $\beta$ & $\theta$ ($\degree$) & Max. separation (mas) \\
    \hline
    Pair 1 & $51339.13^{+0.02}_{-0.02}$ & $15.5^{+0.3}_{-0.3}$ & $7.9^{+0.4}_{-0.3}$ & $0.323^{+0.007}_{-0.006}$ & $0.36^{+0.03}_{-0.02}$  & $28^{+3}_{-3}$ & $19.7^{+0.2}_{-0.2}$\\ 
    \hline
    Pair 2 & $51340.78^{+0.01}_{-0.01}$ & $32.7^{+0.2}_{-0.2}$ & $15.8^{+0.3}_{-0.2}$ & $0.350^{+0.002}_{-0.002}$ & $0.49^{+0.02}_{-0.02}$ & $45^{+2}_{-2}$ & $53.0^{+0.3}_{-0.3}$\\ 
    \hline
    Pair 3 & $50870.12^{+0.02}_{-0.02}$ & $26.7^{+0.2}_{-0.5}$ & $13.7^{+0.2}_{-0.3}$ & $0.320^{+0.004}_{-0.004}$ & $0.43^{+0.02}_{-0.02}$ & $43^{+3}_{-3}$ & $61.2^{+0.6}_{-0.6}$\\ 
    \hline
    \end{tabular}
    \caption{Values derived from the lobe proper motions as shown in Figure \ref{fig:lobes_motion}. The errors are the 16th and 84th percentiles from MCMC posterior distributions. Ejection time is based upon the NE lobe. $\mu_{app}$ and $\mu_{app}$ are the 'true' velocities of the approaching and receding lobes respectively. The maximum separation is of the NE lobe from the core.}
    \label{tab:motions}
\end{table*}

Assuming that the lobes were launched at the same time and moved symmetrically from the core, we calculated for each pair of lobes the ejection angle against the line of sight and intrinsic velocity, using equations from \citet{2006csxs.book..381F}:

\begin{equation}
    \beta_{int}\cos{\theta} = \frac{\mu_{app} - \mu_{rec}}{\mu_{app}+\mu_{rec}}
    \label{eq:beta_cos_theta}
\end{equation}
\begin{equation}
    \tan{\theta} = \frac{2d}{c}\frac{\mu_{app}\mu_{rec}}{\mu_{app}-\mu_{rec}}
    \label{eq:tan_theta}
\end{equation}

The results are displayed in Table \ref{tab:motions}. Our results agreed with those of \citet{2019MNRAS.483.3686M}. We found that the ejection angle of each lobe pair differed, suggesting that the slow jet precesses. \citet{2001ApJ...558..283F}, analysing the same data, found that the angle did not change; this may be because they did not fit for the properties of set of lobes individually but instead averaged across all the data.


\begin{figure}
    \centering
    \includegraphics[width=\linewidth]{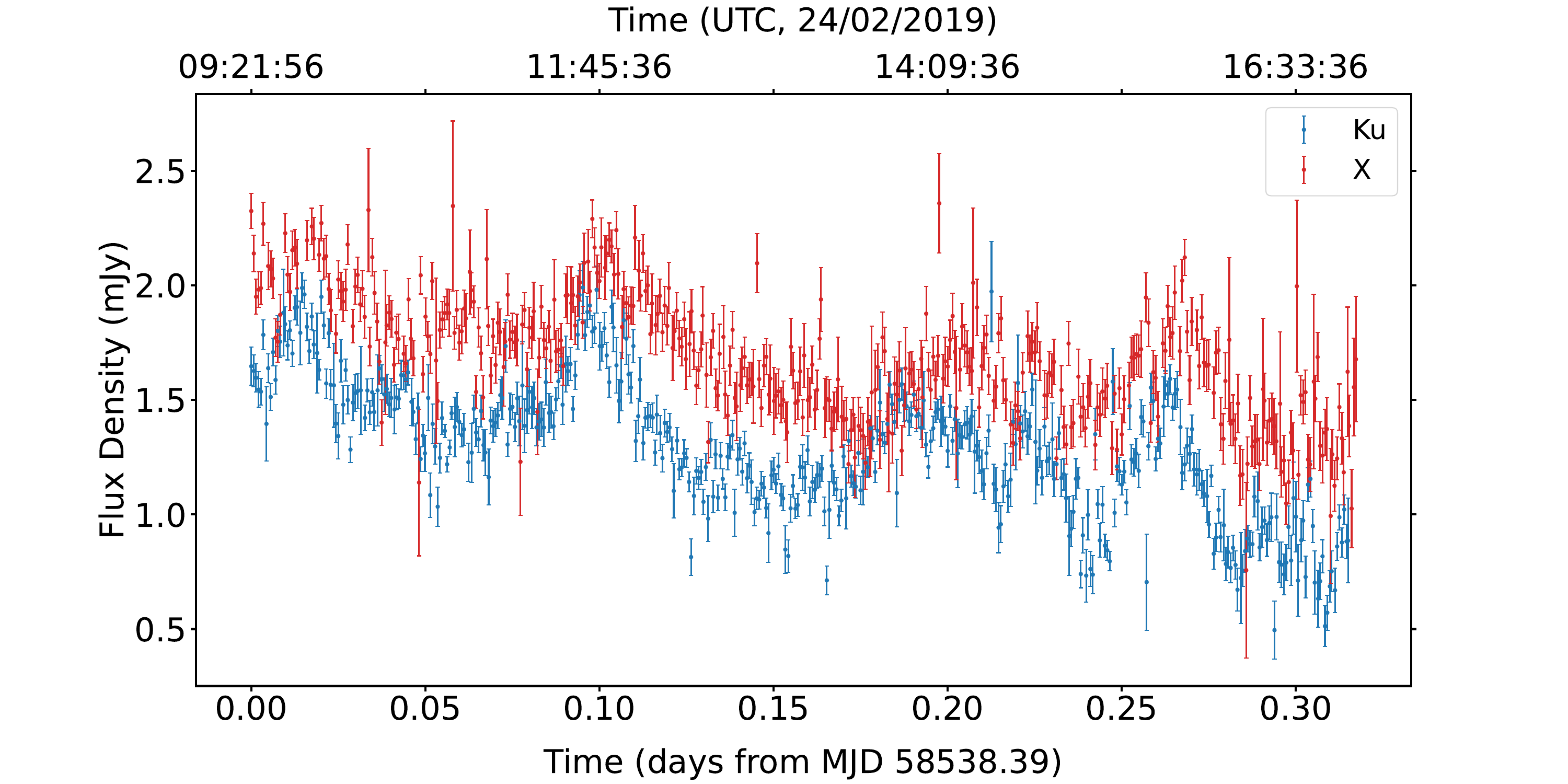}
    \caption{The light curves produced from the VLA data presented in Section \ref{sec:obs} at 10 and 15\,GHz. Each point corresponds to 60\,seconds of data. The light curves show that the emission from Sco X-1 is brighter at lower frequencies but both light curves demonstrate strong flaring behaviour.}
    \label{fig:lightcurve_both}
\end{figure}

\begin{figure}
    \centering
    \includegraphics[width=\linewidth]{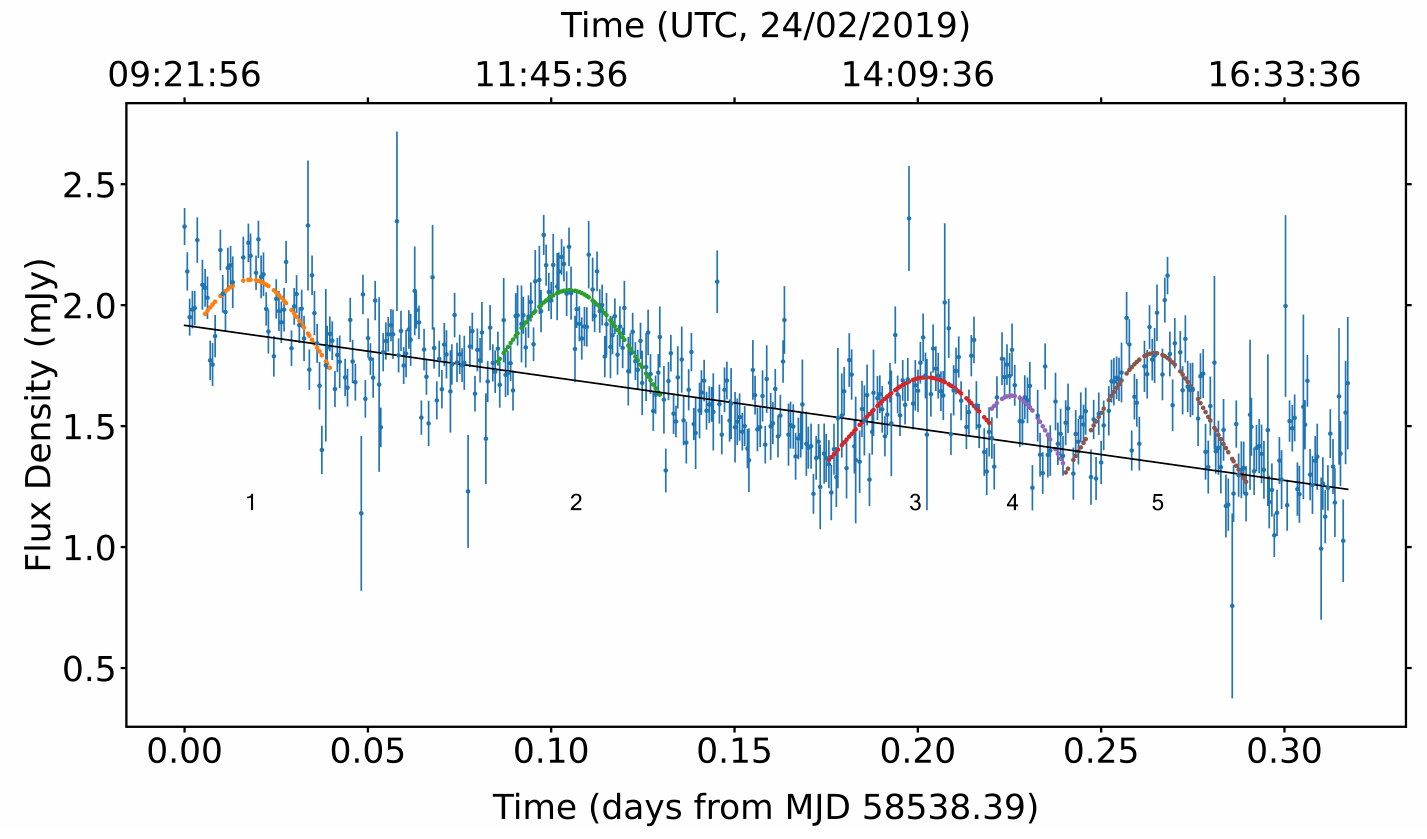}
    \caption{Decomposition of the components in the 10 GHz band. Five flares (labelled 1 to 5) are identified and fitted with broken power laws, and the remaining baseline flux is fitted linearly. Note that the flux after the fifth flare was not included in the linear fit, as the Ku band data suggests there may be a flare, but we were not able to clearly identify it in the X band.}
    \label{fig:lightcurve}
\end{figure}

\subsection{Inferring a URF from spatially-unresolved data}
\label{subsec:URF_vla}

Here, we present how we inferred the presence of URFs from our VLA data. The VLA data (described in Section \ref{sec:obs}) does not spatially resolve the flux from the slow jets (lobes) and the core because the average beam size was $\sim2$ arcseconds which is much larger than the milliarcsecond scales on which the radio components of Sco X-1 have been observed. As a result, our light curves show the total flux from the core and the slow jet, as well as the flares. Figure \ref{fig:lightcurve_both} displays the light curves from our VLA observation, at 10 GHz and 15 GHz. These flares show substantial flux density increases above the baseline in both frequency bands. We note that around 0.05\,days in Figure \ref{fig:lightcurve_both} there is is some low level variability but it is not observed contemporaneously in both bands. As shown in Figure \ref{fig:lightcurve_both}, the flux density at all times (while flaring and between flares) was generally lower in the 15 GHz data compared to 10 GHz, and the flares generally peaked earlier in 15 GHz. For our URF analysis, we considered only the 10 GHz band, as this is closer to the 5 GHz VLBI observations. We make the assumption that the radio spectral properties would not vary drastically between the observed lobes and inferred URF ejections.

With five flares identified, we fitted each with a broken power law, and to the baseline (non-flaring) level of flux that decreased over time, we linear decay. These fits were performed using \textsc{emcee} placing flat priors on all parameters, with 128 walkers and 500 steps, burning the first 100 steps \citep{2013PASP..125..306F}. The linear fit to the baseline flux at 10 GHz produced an estimate of the lowest baseline flux during the observation period as $1.05\pm0.06$ mJy, at MJD 58538.707 (0.317 days from the beginning of the observations).  

As the VLA data did not allow for the radio core and lobes to be spatially resolved, we considered the scenario in which the flares are caused by a URF being launched from the core (causing one flare) and then interacting with the lobes (causing two further flares). We identified every permutation in which a sequence of three flares, all observed at 10\,GHz, might be explained by a URF, by assigning them to the core, approaching lobe and receding lobe (in that order). We found 10 sequences of flares (see Table \ref{tab:results}, which uses the labels in the lower panel of Figure \ref{fig:lightcurve}), which we considered equally in the subsequent analysis. For each sequence, we calculated the time intervals between flares as measured using the peak times derived from the broken power law fits as a proxy for the time between the launch from the core to the interaction in the lobes. Assuming that the URF was ejected at $45\pm2\degree$ to the line of sight (the largest inclination inferred in Section \ref{subsec:slowjet}, and discussed further in the next paragraph), and reached a maximum separation of 53 mas (i.e. that it followed the same path, for the same distance, as lobe pair 2, we combined the observed time between flares with Equations \ref{eq:beta_cos_theta} and \ref{eq:tan_theta} to infer the speed of the URF.

\begin{figure}
    \centering
    \includegraphics[width=0.8\linewidth]{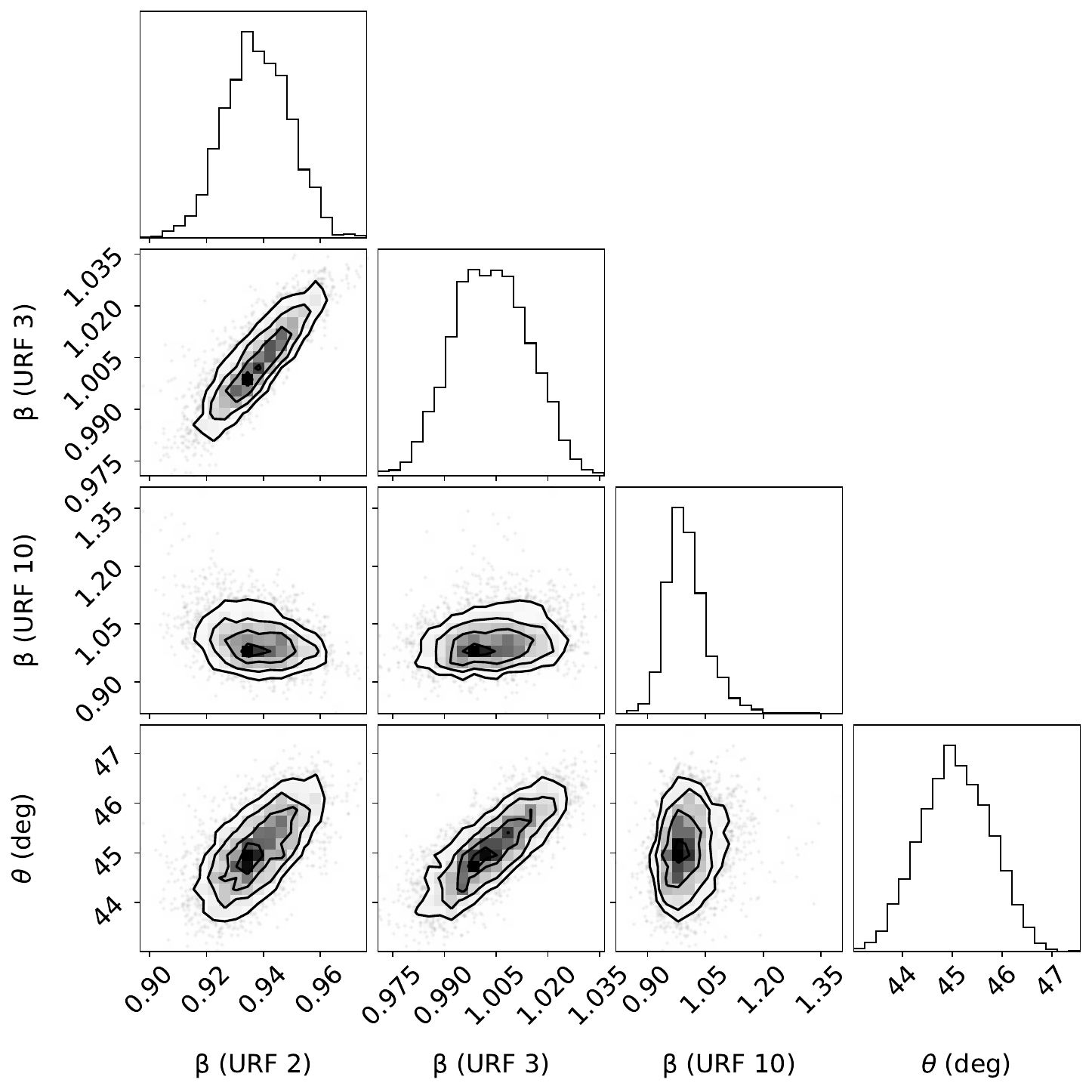}
    \caption{Corner plots of the derived posterior distribution for the three fastest URF candidates. The plots show how inferred velocity depends on the angle against the plane of the sky of the jet. For URFs 2 and 3, the inferred velocity depends strongly on angle.}
    \label{fig:cornerplots}
\end{figure}

Table \ref{tab:results} displays the inferred Lorentz factors with errors caulcated from the posteriors of our \textsc{mcmc} fitting. The majority of flare sequences, with inferred velocities between 0.5c and 0.8c, did not suggest the presence of a URF. One sequence, with $\beta=0.89$, $\Gamma=2.2$, did not quite reach the ultra-relativistic regime. Three sequences allowed for ultra-relativistic or even superluminal URFs to be inferred: the second, with $\beta=0.94$, $\Gamma=2.9$; the third, with $\beta=1.00$, $\Gamma>8.1$; and the tenth, with $\beta=1.00$, $\Gamma>3.4$. The uncertainties were larger for sequences that involved the second, third and particularly fourth flares, as the uncertainty in determining their peak times was higher. The results were highly sensitive to angle, as shown in Figure \ref{fig:cornerplots}, the bottom row of which displays the dependence of the \textsc{mcmc}-inferred velocity on the adopted URF angle. In the bottom-left panel, for example, the inferred velocity of URF series 2 increases by about $\beta=0.02$ as the inclination angle increases by $1\degree$. If 43$\degree$ was adopted as the inclination angle of the URF, as in the third lobe pair of the slow jet, then none of the URFs reached superluminal inferred velocities, but series 2 and 10 were still ultra-relativistic, with $\Gamma=2.87$ and $\Gamma=2.69$ respectively. If 28$\degree$ was adopted, as in the first lobe pair of the slow jet, then the URFs could only reach around 0.8c. 

\begin{figure}
    \centering
    \includegraphics[width=\linewidth]{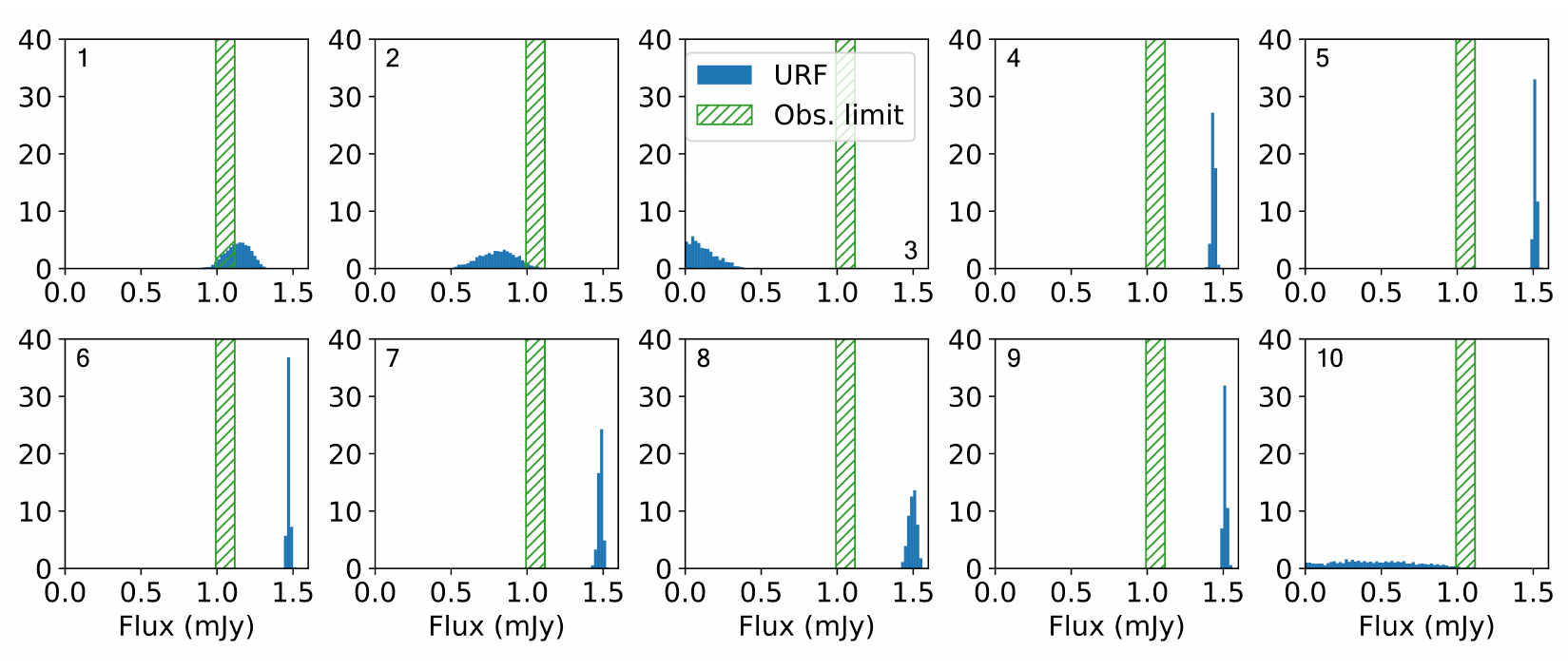}
    \caption{Inferred observed flux of each potential URF. Results are displayed as histograms of probability density, calculated from the propagation of \textsc{mcmc} posterior distributions. Each bin covers an interval of 0.02 mJy. The green `observational limit' bar represents the lowest level of flux observed from the X-band 2019 observations. If the blue histogram is to the left of the green bar, the URF would be unobserved. The green bar is also derived from \textsc{mcmc} posterior distributions and represents the 16th to 84th percentile range. Each panel has the series number (from Table \ref{tab:results}) labelled.}
    \label{fig:urf_obsflux}
\end{figure}

\subsection{Effects of Relativistic Beaming}
\label{subsec:beaming}

We next consider the effects of relativistic beaming, from the relativistic motion of the URF, on the observed emission of the URF, which we now assume is intrinsically luminous. Despite it being luminous, we only detect the URF when it is launched and interacts with the lobes. As we did not directly detect a URF in our data, we can assume that an inferred URF would relativistically beam its flux below the lowest baseline level of flux observed, that was fitted for in Figure \ref{fig:lightcurve}. The difference between observed flux $F_{\textrm{obs}}$ and the emitted flux $F_{\textrm{em}}$ is given by:
\begin{equation}
    F_{\textrm{obs}} = F_{\textrm{em}} \delta^k
    \label{eq:F_obs}
\end{equation}

where
\begin{equation}
    \delta = \Gamma^{-1}(1\mp\beta_{int}\cos{\theta})
    \label{eq:delta}
\end{equation}
and $k=3$ for flux dominated by a single evolving event such as a lobe \citep{2006csxs.book..381F}.

Our analysis relies on the assumption that the rest-frame flux density of the URF was the same as that of the non-flaring slow jet/lobe that we analysed in Section \ref{subsec:slowjet}, i.e. that they have the same intrinsic properties. We use the VLA observations from June 1999 (shown in Figure \ref{fig:lobes_motion}) as a means to calculate the intrinsic jet properties. The VLA observations found that the NE lobe had a flux density of 1.35\,mJy at 5\,GHz \citep[][]{2001ApJ...558..283F}. The flux density of the SW component was hard to determine as there were few datapoints. Using equations \ref{eq:F_obs} and \ref{eq:delta}, where $F_{\textrm{obs}} = 1.35$\,mJy and $\beta$, $\Gamma$ and $\theta$ as from the second row of Table \ref{tab:motions}, we calculated the rest frame flux density of the lobe and therefore the URF to be $0.76\pm0.01$ mJy.

We then used the rest frame flux density to determine the observed flux of the URF for each flare sequence, using the different inferred velocities in Table \ref{tab:results} along with Equations \ref{eq:F_obs} and \ref{eq:delta}. Where the inferred URF velocity was above $c$, and so $\delta$ in Equation 4 was infinite, the observed flux was assumed to be 0. For the URF to be unobserved, its flux would have to fall below the `observational limit' of $1.05\pm0.06$ mJy (the lowest non-flaring level of flux density from the 10 GHz VLA observations in Figure \ref{fig:lightcurve}). Figure \ref{fig:urf_obsflux} shows the posterior distributions for our derived observed fluxes compared to this limit for each inferred URF. As expected, the three sequences with ultra-relativistic inferred velocities were also those which would not be observed. Due to the steep dependence of Doppler factor on velocity at values close to c, the URF candidates with highest velocities had greater uncertainty in their inferred observed flux.

\def\arraystretch{1.2}
\begin{table}
 \caption{Full results for values inferred for the URFs. Errors are the 16th and 84th percentile of the derived posterior distribution. Lower limits are quoted for $\Gamma$ where the velocity exceeded $c$ and so $\Gamma$ tended to infinity. The numbers in the `Flare sequence' column refer to the flares labelled in Figure \ref{fig:lightcurve}. Three of the possible flare permutations are allowed: 2, 3 and 10. }
 \label{tab:results}
 \begin{tabular}{lccc}
  \hline
  Series No. & Flare sequence & Velocity ($\Gamma$) & Inferred flux density (mJy) \\
  \hline
  1 & 1-2-3 & $2.2^{+0.2}_{-0.1}$ & $1.14^{+0.08}_{-0.08}$\\
  2 & 1-2-4 & $2.9^{+0.5}_{-0.2}$ & $0.8^{+0.1}_{-0.1}$\\
  3 & 1-2-5 & $>8.1$ & $0.0^{+0.1}_{-0.0}$\\
  4 & 1-3-4 & $1.21^{+0.01}_{-0.02}$ & $1.44^{+0.01}_{-0.01}$\\
  5 & 1-3-5 & $1.33^{+0.02}_{-0.01}$ & $1.51^{+0.01}_{-0.01}$\\
  6 & 1-4-5 & $1.25^{+0.02}_{-0.01}$ & $1.470^{+0.009}_{-0.008}$\\
  7 & 2-3-4 & $1.27^{+0.04}_{-0.02}$ & $1.48^{+0.01}_{-0.02}$\\
  8 & 2-3-5 & $1.56^{+0.07}_{-0.03}$ & $1.50^{+0.03}_{-0.03}$\\
  9 & 2-4-5 & $1.33^{+0.04}_{-0.02}$ & $1.51^{+0.01}_{-0.01}$\\
  10 & 3-4-5 & $>3.4$ & $0.04^{+0.06}_{-0.4}$\\
  \hline
 \end{tabular}
\end{table}

\section{Discussion}\label{sec:Discussion}

We have shown that radio flares from the Sco X-1 system are capable of supporting flows travelling at $\Gamma>2$. Flares from Sco X-1 have previously been shown to occur in a correlated manner, suggesting the presence of a URF travelling down the path of the `slow' jet. Under the assumption that URFs and slow jets share many intrinsic properties, we show that a URF with equal rest-frame flux density to the slow jet would undergo sufficient relativistic boosting out of our line of sight that its flux would be `hidden' by the flux from the observed slow jet. Three combinations of flares, numbers 2, 3 and 10 as presented in Table \ref{tab:results}, have a high $\Gamma$ and are estimated to beam their flux well below the lowest level observed by the VLA. Flare combinations 2, 3 and 10 have $\Gamma$ values or limits of $2.9^{+0.5}_{-0.2}$, $>8.1$ and $>3.4$, respectively. Fast jets have been observed in other X-ray binaries \citep[e.g. GRS 1915$+$105, MAXI J1820$+$070][]{1999MNRAS.304..865F,2020NatAs...4..697B}. Our results suggest similar jets could be produced in the Sco X-1 system and be the cause of the flaring observed in the radio lobes, despite no observations of fast jets in Sco X-1. 

We have made a number of assumptions about the similarity of the URF to the slow jet, and we note that while the URF remains unseen few scenarios can be ruled out concerning its true nature. The assumption that a re-flaring event requires the URF to flow along the same axis as the slow jet assumes a relatively small opening angle for a URF, whereas a flow with a wide opening angle may interact with jet ejecta even if launched at a different angle. Changing the assumed launch angle of the URF can have drastic consequences for the inferred velocity (as shown in Figure \ref{fig:cornerplots}), and therefore the ability of the URF to beam its flux out of the line of sight. The discussion of `hidden' URFs may be moot if URFs are not intrinsically luminous, or invalid if URFs are more highly luminous than expected. However, we argue that the existing indirect evidence, concerning the radio spectral properties of Sco X-1 during slow jet and URF launches, does not suggest that the URF is intrinsically different to the slow jet.

\citet{Fender_Motta_2025} reframe the observations of proper motion-resolved radio jets such that `slow' and `fast' jets are launched through different mechanisms. `Slow' jets are those with $\beta\Gamma$ values below 2, and `fast' jets are those with $\beta\Gamma$ above 2. They also demonstrate that fast jets are always launched in the same direction, whereas slower jets can precess. We find that the three high-$\Gamma$ flare combinations have $\beta\Gamma$ values above 2, consistent with the `fast' jet regime. Additionally, we find that the three pairs of observed radio lobes presented in Table \ref{tab:motions} all have $\beta\Gamma$ values far below $\beta\Gamma = 2$, consistent with `slow' jets. 
It has been shown here, by \citet{2001ApJ...553L..27F} and \citet{2019MNRAS.483.3686M} that the slow jet-driven lobes from Sco X-1 have at least two distinct angles to the line of sight: $\sim28\degree$ and  $\sim44\degree$. Whilst we have concluded that the flares observed by the VLA correspond to either the launch of fast jets or interaction between them and the slow jets, we do not have sufficient angular resolution to spatially separate or measure the angle to the line of sight of the jets, to determine if there is precession as was previously observed. Assuming that the URF is similar in nature to the slow jet  - having a small opening angle and propagating along a specific direction - we do not expect to observe re-flaring events when the slow jets are launched at smaller angles to the line of sight because the URFs, which are launched along a fixed axis, will not interact with them. However, we note that such conclusions can be drawn in future works examining contemporaneous VLBI and lower angular resolution campaigns (Motta et al \textit{in prep}). Such results will be able to break the degeneracy between variability as a result of jet-ISM interaction and slow-fast jet interaction, and determine whether the scenario we have put forward here can explain the radio data from Sco X-1.

\section{Conclusions}\label{sec:Conclusion}

We used archival radio data to determine the proper and intrinsic motions of the spatially resolved radio lobes of Sco X-1. We found that pairs of radio lobes were emitted at speeds of around 0.4c, and at inclination angles between 28-45$\degree$. We analysed new, low angular resolution radio data from the VLA showing the variation of radio emission from the whole Sco X-1 system with time. 

While we were unable to determine the source of each flare as the lobes were not spatially resolved, we found that several sequences of flares could plausibly be explained by a fast jet (or `URF') causing a flare as it is emitted from the core, travelling down the pathway of the observed radio jet at 45$\degree$ (angle to line of sight), reaching the lobes at their maximum separation (53 mas for the NE lobe), and re-brightening each lobe. The inferred velocities of such flows could reach the speed of light. Assuming that the rest-frame flux density of the flow is the same as that of the slow jet, 0.76 mJy, the flux from the fast jets would be relativistically beamed out of our line of sight to a level well below the observational limit, and would therefore go unseen. 

In future work, observations with high-quality spatial and temporal resolution would be able to confirm or deny whether this explanation for Sco X-1 is plausible. 

\section*{Acknowledgements}

The National Radio Astronomy Observatory is a facility of the National Science Foundation operated under cooperative agreement by Associated Universities, Inc.

L.~R. acknowledges support from the Trottier Space Institute Fellowship and from the Canada Excellence Research Chair in Transient Astrophysics (CERC-2022-00009). A.~J.~C acknowledges support from the Oxford Hintze Centre for Astrophysical Surveys which is funded through generous support from the Hintze Family Charitable Foundation.

\section*{Data Availability}

The data presented in this Letter are provided in a machine readable table in the supplementary material.



\bibliographystyle{mnras}
\bibliography{bib} 






\bsp	
\label{lastpage}
\end{document}